\documentclass[aps,prl,reprint,superscriptaddress,noeprint]{revtex4-1}

\usepackage{amssymb}
\usepackage{graphicx}
\usepackage{color}

\newcommand{\araa}{Annu. Rev. Astron. Astrophys.}
\newcommand{\aap}{Astron. Astrophys.}
\newcommand{\aapr}{Astron. Astrophys. Rev.}
\newcommand{\aj}{Astron. J.}
\newcommand{\apjl}{Astrophys. J. Lett.}
\newcommand{\baas}{Bulletin American Astron. Soc.}
\newcommand{\mnras}{Mon. Not. R. Astron. Soc.}
\newcommand{\pasa}{Publ. Astron. Soc. Australia}
\newcommand{\physrep}{Phys. Rep.}

\newcommand{\be}{\begin{equation}}
\newcommand{\ee}{\end{equation}}
\newcommand{\ba}{\begin{eqnarray}}
\newcommand{\ea}{\end{eqnarray}}
\newcommand{\bd}{\begin{displaymath}}
\newcommand{\ed}{\end{displaymath}}
\newcommand{\bano}{\begin{eqnarray*}}
\newcommand{\eano}{\end{eqnarray*}}
\newcommand{\lp}{\left(}
\newcommand{\rp}{\right)}

\newcommand{\nus}{\nu_\mathrm{s}}
\newcommand{\nugw}{\nu_\mathrm{gw}}
\newcommand{\taugw}{\tau_\mathrm{gw}}
\newcommand{\Msun}{M_{\rm Sun}}
\newcommand{\hamp}{h_0}

\begin{document}

\title{Gravitational waves from transient neutron star f-mode oscillations}

\author{Wynn C. G. Ho}
\email[]{wynnho@slac.stanford.edu}
\affiliation{Department of Physics and Astronomy, Haverford College, 370 Lancaster Avenue, Haverford, PA, 19041, USA}
\affiliation{Mathematical Sciences and STAG Research Centre, University of Southampton, Southampton SO17 1BJ, UK}
\author{D.I. Jones}
\author{Nils Andersson}
\affiliation{Mathematical Sciences and STAG Research Centre, University of Southampton, Southampton SO17 1BJ, UK}
\author{Crist\'obal M. Espinoza}
\affiliation{Departamento de F\'isica, Universidad de Santiago de Chile, Avenida Ecuador 3493, 9170124 Estaci\'on Central, Santiago, Chile}

\date{\today}

\begin{abstract}
During their most recent observing run, LIGO/Virgo reported the
gravitational wave (GW) transient S191110af, a burst signal
at a frequency of 1.78~kHz that lasted for 0.104~s.
While this signal was later deemed non-astrophysical, genuine
detections of uncertain origin will occur in the future.
Here we study the potential for detecting GWs from neutron star
fluid oscillations, which have mode frequency
and duration matching those of S191110af and which can be used to
constrain the equation of state of nuclear matter.
Assuming that such transient oscillations can be excited to energies
typical of a pulsar glitch,
we use measured properties of known glitching pulsars to estimate
the amplitude of GWs produced by such events.
We find that current GW detectors may observe nearby pulsars undergoing
large events
with energy similar to Vela pulsar glitch energies, while next generation
detectors could observe a significant number of events.
Finally, we show that it is possible to distinguish between GWs
produced by rapidly rotating and slowly rotating pulsars from
the imprint of rotation on the f-mode frequency.
\end{abstract}

\maketitle

\paragraph{Introduction\label{sec:intro}}
The new era of gravitational wave (GW) astronomy began with the
detection of binary black hole and binary neutron star (NS)
mergers in the last several years by the advanced GW detectors of
Laser Interferometer Gravitational-wave Observatory (LIGO)
and Virgo \cite{abbottetal16,abbottetal17}.
Searches are ongoing for more black hole and NS mergers,
as well as for NS-black hole mergers and other GW sources such as
transient signals associated with supernovae and fast radio bursts (FRBs).
Like in traditional electromagnetic astronomy, there may be occasions
when ``rare'' GW signals are detected whose properties are not
well-understood or modeled at the time of discovery.
For example, the phenomenon of FRBs was not equivocally known to have
an astrophysical origin when the first one was found in 2007
\cite{lorimeretal07}, and the astrophysical nature of FRBs is still not
known even after more than one hundred events have been detected
\cite{petroffetal16,cordeschatterjee19,petroffetal19,plattsetal19}.

In the GW regime, a somewhat analogous signal to FRBs was reported recently.
The GW transient candidate S191110af was detected on 2019 November 10
by LIGO/Virgo and consists of a signal at 1.78~kHz that lasted for 0.104~s
\cite{chatterjee19}.
Follow-up analysis over the next few days identified instrumental
artifacts in the data, which led to retraction of S191110af as a
genuine astrophysical signal \cite{chatterjee19b}.
In the intervening time, it was pointed out
that the frequency and burst duration of S191110af
are consistent with the fundamental stellar oscillation mode (f-mode)
of a NS of mass $M=1.25\,\Msun$ and radius $R=13.3\mbox{ km}$
\cite{anderssonkokkotas98,kaplanetal19}.
In addition, the results of \cite{kokkotasetal01,keerjones15} were
used to estimate that a f-mode could produce a GW signal-to-noise
ratio (SNR) $\sim 10$ \cite{kaplanetal19}.
GW-producing f-mode oscillations can be triggered by transient
events internal to the NS
\cite{anderssoncomer01,vaneysdenmelatos08,sideryetal10,keerjones15},
such as a sudden phase transition, magnetic field reorganization,
or pulsar glitch
(a sudden change $\Delta\nus$ of NS spin rate $\nus$ due to starquakes
or more likely angular momentum exchange between normal and superfluid
components within the star; \cite{andersonitoh75,haskellmelatos15}).
However, evidence of glitching pulsars that could be responsible for
S191110af was not found \cite{kaplanetal19}.
The effectiveness of different proposed mechanisms is also unclear
since their energy may not be released as GWs.

GWs associated with the f-mode have been of great interest because
the mode frequency depends on the dynamical timescale of NSs
and hence is a probe of NS density, mass, and radius
\cite{anderssonkokkotas96,anderssonkokkotas98,kokkotasetal01}.
A f-mode GW signal can appear in newborn NSs \cite{ferrarietal03},
magnetars \cite{abbottetal19}, and NS mergers, during both pre-merger
\cite{shibata94,kokkotasschafer95,steinhoffetal16,anderssonho18,vicklai19}
and
post-merger \cite{xingetal94,stergioulasetal11,bausweinjanka12,abbottetal19b}
phases.
Here we expand on the brief analysis of \cite{kaplanetal19}
and study detectability of GWs produced by a f-mode in rotating NSs,
assuming that the mode is excited to a level corresponding to the
energy associated with typical pulsar glitches \cite{anderssoncomer01}.
This is sensible because we know that pulsars exhibit transients
at this level, even though there is no established connection between
mode excitation and observed glitches.
The advantage is that, by considering normal isolated NSs that
are well-studied, we have a well-defined source population with
known properties, and we avoid the uncertainties of speculating
on and modeling unknown sources.
Moreover, this allows us to consider a question that may become
relevant in the future: How do we distinguish astrophysical
transients from detector noise if both are associated with
exponentially damped sinusoidal signals?

\paragraph{Model for GW source\label{sec:model}}
Consider a stellar oscillation with frequency $\nugw$ that is induced
at time $t=0$ and damps on a timescale $\taugw$.
Following \cite{echeverria89,finn92},
the GW amplitude from such an oscillation is then zero for $t<0$ and
\be
h(t) = \hamp\,e^{-t/\taugw}\sin(2\pi\nugw t)\qquad\mbox{ for $t>0$}.
\ee
The peak amplitude $\hamp$ can be determined by first noting that
the GW luminosity of a source at distance $d$ is \cite{owen10}
\be
\frac{dE_{\rm gw}}{dt} = \frac{c^3d^2}{10G}
 \lp2\pi\nugw\hamp e^{-t/\taugw}\rp^2. \label{eq:lumin}
\ee
We then integrate equation~(\ref{eq:lumin}) over $0<t<\infty$ to obtain
the total GW energy emitted $E_{\rm gw}$ and solve for $\hamp$ to find
\ba
\hamp&=&\frac{1}{\pi d\nugw}\lp\frac{5G}{c^3}\frac{E_{\rm gw}}{\taugw}\rp^{1/2}
 = 4.85\times 10^{-17}\lp\frac{1\mbox{ kpc}}{d}\rp
\nonumber\\
&& \times \lp\frac{E_{\rm gw}}{\Msun c^2}\rp^{1/2}
\lp\frac{1\mbox{ kHz}}{\nugw}\rp\lp\frac{0.1\mbox{ s}}{\taugw}\rp^{1/2}.
 \label{eq:hamp1}
\ea

Now consider the oscillation mode is excited to a level corresponding
to a pulsar glitch, such that the GW energy $E_{\rm gw}$ is supplied
by the energy of the glitch
\ba
E_{\rm glitch} &=& 4\pi^2I\nus\Delta\nus
\nonumber\\
&=& 3.95\times 10^{40}\mbox{ erg }
 \lp\frac{\nus}{10\mbox{ Hz}}\rp \lp\frac{\Delta\nus}{10^{-7}\mbox{ Hz}}\rp,
\label{eq:eglitch}
\ea
where NS moment of inertia $I\sim 10^{45}\mbox{ g cm$^{2}$}$.
Substituting equation~(\ref{eq:eglitch}) into equation~(\ref{eq:hamp1}),
the peak GW amplitude is
\ba
\hamp &=& 7.21\times 10^{-24} \lp\frac{1\mbox{ kpc}}{d}\rp
 \lp\frac{\nus}{10\mbox{ Hz}}\rp^{1/2}
 \lp\frac{\Delta\nus}{10^{-7}\mbox{ Hz}}\rp^{1/2}
\nonumber\\
&& \times
 \lp\frac{1\mbox{ kHz}}{\nugw}\rp\lp\frac{0.1\mbox{ s}}{\taugw}\rp^{1/2}.
\label{eq:hamp}
\ea
Thus for a given oscillation mode frequency $\nugw$ and damping
time $\taugw$, the peak GW amplitude $\hamp$ depends
on distance $d$ to the pulsar, the pulsar spin frequency $\nus$,
and glitch size $\Delta\nus$.
In this calculation,
we consider the f-mode oscillation to be efficient at extracting
energy at the level of glitches and driving the emission of GWs.
In reality, such a process is likely to be at least somewhat inefficient.
However, a factor of, e.g., ten lower energy that is converted
to GWs (from a glitch or other process) only reduces the GW
amplitude by a factor of three since $\hamp\propto\sqrt{E_{\rm gw}}$.
On the other hand, g-modes are known to be significantly less
efficient than f-modes at producing GWs
\cite{lai99,ferrarietal03,krugeretal15}.

\paragraph{Pulsar and glitch distributions\label{sec:glitch}}
For our nominal GW sources, we use 552 glitches from 188 pulsars in the
Jodrell Bank Glitch Catalogue
\cite{espinozaetal11}\footnote{http://www.jb.man.ac.uk/pulsar/glitches.html}.
The Jodrell Bank Glitch Catalogue lists the relative spin frequency
change $\Delta\nus/\nus$ for each detected glitch.
We use the ATNF Pulsar Catalogue \cite{manchesteretal05} to
supplement the glitch data with each pulsar's spin frequency $\nus$,
distance $d$, and sky position.
Note that the default distance in the ATNF Pulsar Catalogue is derived
from each pulsar's dispersion measure \cite{yaoetal17},
although in some cases an independent distance is known.
Since we are not focused on most individual pulsars but on the
overall population, distance errors are not important.

%-------------------------------------------------
\begin{figure}[!tb]
\includegraphics[width=0.45\textwidth]{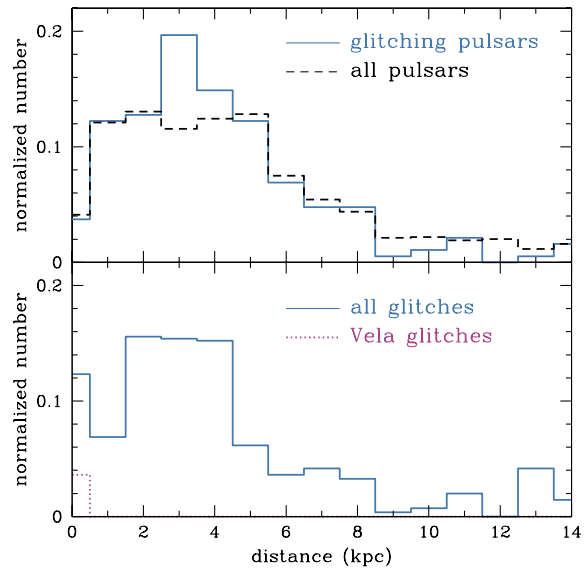}
\vspace{-1em}
\caption{
Top: Normalized distributions of distance for all pulsars (dashed line)
and for all glitching pulsars (solid line).
Bottom: Normalized distribution of distance for all glitches (solid line).
Dotted line is the distribution for Vela glitches.
\label{fig:psrdist}}
\vspace{-0.5em}
\end{figure}
%-------------------------------------------------

%-------------------------------------------------
\begin{figure}[!tb]
\includegraphics[width=0.45\textwidth]{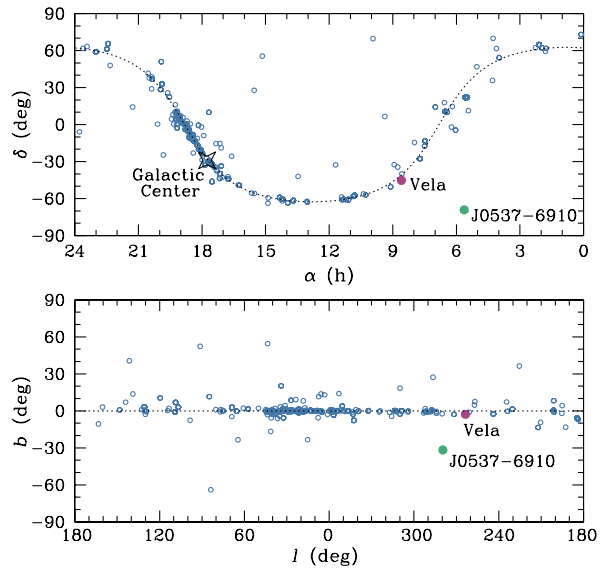}
\vspace{-1em}
\caption{Position of glitching pulsars in right ascension $\alpha$
and declination $\delta$ (top) and Galactic longitude $l$ and latitude
$b$ (bottom).
The Galactic plane is denoted by dotted lines,
and positions of the Galactic Center and pulsars Vela and PSR~J0537$-$6910
are labeled.
\label{fig:psrsky}}
\vspace{-0.5em}
\end{figure}
%-------------------------------------------------

The top panel of
Figure~\ref{fig:psrdist} shows the normalized distance distributions
of all $\sim 2700$ known pulsars (with a distance) in the ATNF Pulsar
Catalogue and 188 glitching pulsars in the Jodrell Bank Glitch Catalogue.
The bottom panel shows the distance distribution of the 552 glitches.
We see that a majority of known pulsars and glitching pulsars are
at distances $d<6\mbox{ kpc}$.
This is due in large part to observational selection effects.
The Vela pulsar, at a distance of 287~pc \cite{dodsonetal03},
contributes significantly to the very nearby glitch population.
Meanwhile, PSR~J0537$-$6910, also known as the Big Glitcher
\cite{marshalletal04},
is in the Large Magellanic Cloud at a distance of 50~kpc and
thus is not shown in Figure~\ref{fig:psrdist}.
We also need to keep in mind that the population of ``seismically active''
NSs which emit GWs could be dominated by objects that have not yet
been detected.

Figure~\ref{fig:psrsky} shows the position of each glitching pulsar.
Glitching pulsars are clearly clustered in the Galactic plane,
as expected for relatively young pulsars.
Because source localization by only GW detectors is generally poor,
we may not be able to determine definitively whether an individual
GW burst originates from a source in the Galactic plane.
However, such a determination may be possible for a population of
burst sources (such as the glitching pulsars) if they all contain localization
regions that overlap with the Galactic plane or even cluster near
the Galactic Center \cite{kimdavies18}.

%-------------------------------------------------
\begin{figure}[!tb]
\includegraphics[width=0.45\textwidth]{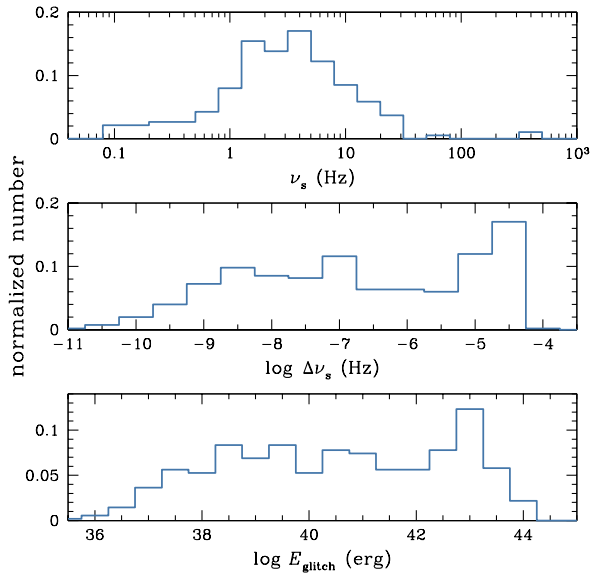}
\vspace{-1em}
\caption{
Normalized distributions of spin frequency $\nus$ (top) for all
188 glitching pulsars and glitch size $\Delta\nus$ (middle) and
glitch energy $E_{\rm glitch}$ (bottom) for all 552 glitches.
\label{fig:glitch}}
\vspace{-0.5em}
\end{figure}
%-------------------------------------------------

The top panel of
Figure~\ref{fig:glitch} shows the distribution of spin frequency
$\nus$ for the 188 glitching pulsars,
and the middle and bottom panels show the distributions of glitch
size $\Delta\nus$ and glitch energy $E_{\rm glitch}$, respectively,
for the 552 glitches.
Most glitching pulsars have relatively low spin frequencies,
i.e., $\nus\approx 1-30\mbox{ Hz}$.
Glitch size has a broad range $\Delta\nus\sim 10^{-9}-10^{-5}\mbox{ Hz}$
\cite{espinozaetal11,yuetal13,fuentesetal17},
which leads to a broad range of glitch energies
$E_{\rm glitch}\sim 10^{37}-10^{44}\mbox{ erg}\sim 10^{-17}-10^{-10}\Msun c^2$.

\paragraph{Results\label{sec:results}}
With the known properties of our model source population described above,
we compute the amplitude of GWs emitted from a damped f-mode triggered by
the energy equivalent to a pulsar glitch [see equation~(\ref{eq:hamp})].
First, we must determine the f-mode frequency and damping time.
Early works
\cite{anderssonkokkotas96,anderssonkokkotas98,kokkotasetal01}
show that $\nugw$ and $\taugw$ are related to NS mass $M$ and
radius $R$ in a way that is approximately independent of nuclear
equation-of-state (EOS).
Subsequent work verified these relations \cite{donevaetal13} and find
alternative relations that depend on $M$ and moment of inertia $I$
\cite{chirentietal15,donevakokkotas15}.

%-------------------------------------------------
\begin{figure}[!tb]
\includegraphics[width=0.45\textwidth]{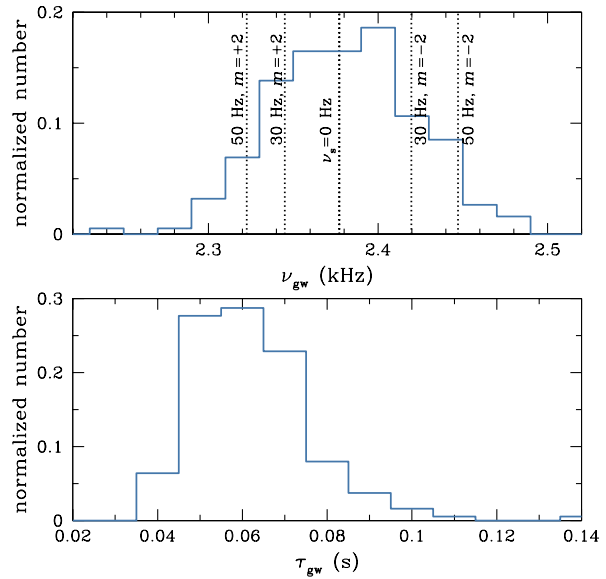}
\vspace{-1em}
\caption{
Normalized distributions of f-mode frequency $\nugw$ (top)
and damping time $\taugw$ (bottom), derived using a Gaussian mass
distribution peaked at $M=1.4\Msun$ with $0.15\Msun$ width,
the BSk24 EOS for the radius, and non-rotating $\nugw(M,R)$ and
$\taugw(M,R)$ relations of \cite{donevaetal13}.
Vertical dotted lines indicate (inertial) frame $\nugw$ for labeled
spin frequency and oscillation mode order $m$ and using spin corrections
of \cite{donevaetal13} with $\nu_{\rm K}=1\mbox{ kHz}$.
\label{fig:ftau}}
\vspace{-0.5em}
\end{figure}
%-------------------------------------------------

%-------------------------------------------------
\begin{figure*}[!tb]
\includegraphics[width=0.7\textwidth]{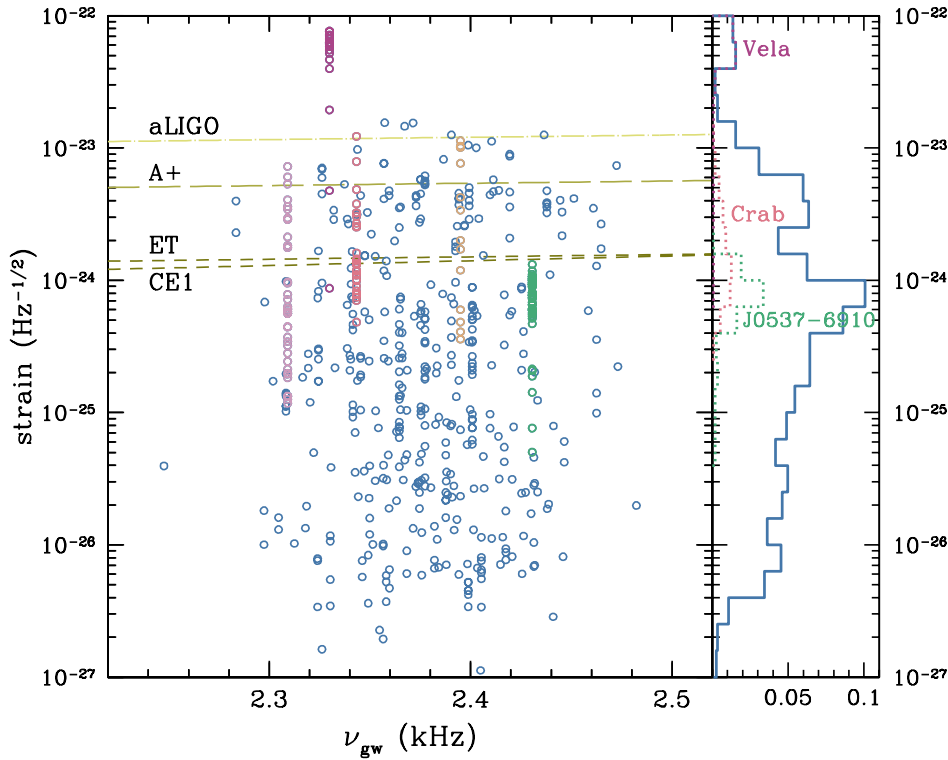}
\vspace{-1em}
\caption{
GW spectrum.
Each circle denotes the peak strain (=$\hamp\sqrt{\taugw}$) of a burst
of GWs emitted at $\nugw$ from a damped f-mode oscillation excited
to a level equivalent to the energy of
one of the 552 known glitches in the Jodrell Bank Glitch Catalogue.
Nearly horizontal dashed lines are sensitivities of
Advanced LIGO (aLIGO), A+, Einstein Telescope (ET), and
Cosmic Explorer (CE1).
Right panel: Normalized strain distribution.
Bursts attributed to glitches of PSR~B1737$-$30, Vela and Crab pulsars,
PSR~J0205+6449, and PSR~J0537$-$6910 are highlighted (from left to right).
\label{fig:strain}}
\vspace{-0.5em}
\end{figure*}
%-------------------------------------------------

For simplicity, we use the f-mode frequency and damping time relations
of \cite{donevaetal13} to $M$ and $R$ in the non-rotating limit (see below).
We randomly assign a mass to each of the 188 glitching pulsars,
where $M$ is drawn from a Gaussian distribution centered at
$M=1.4\Msun$ with a width of $0.15\Msun$.
The radius is then determined from the mass using the BSk24 EOS
\cite{pearsonetal18},
which is a modern nuclear EOS that we choose simply as an example.
The BSk24 EOS generates NSs whose mass and radius satisfy the
$M$--$R$ constraints from {\it NICER} \cite{milleretal19,rileyetal19}
and produces a maximum NS mass that exceeds the highest observed NS mass
\cite{cromartieetal20}.
Figure~\ref{fig:ftau} shows the resulting distributions of f-mode
frequency and damping time.
We note that the peak f-mode frequency and damping time for the
somewhat softer APR EOS \cite{akmaletal98} are at 2.5~kHz and
0.05~s, respectively.
Since $\nugw\propto\sqrt{M/R^3}$ and $\taugw\propto R^4/M^3$
\cite{anderssonkokkotas96,anderssonkokkotas98,donevaetal13},
our assumed mass distribution with width
$\sim 10\%$ produces f-mode and damping time distributions
with width $\sim 5\%$ and $\sim 30\%$, respectively.
On the other hand, there is generally not much difference
between radii of different masses around $1.4\Msun$ for
a given EOS, e.g., the radius differs by $<1\%$ in the mass
range $M=1.1$--$1.7\Msun$ for BSk24 ($<3\%$ for APR) .
Therefore radius variations do not contribute significantly
to variations of $\nugw$ and $\taugw$.

The leading order spin corrections to $\nugw$ are
$\approx (0.2-0.4)(\nus/\nu_{\rm K})$ \cite{donevaetal13},
where Kepler frequency
$\nu_{\rm K}\approx\sqrt{GM/R^3}/3\pi\sim 1\mbox{ kHz}$
\cite{shapiroteukolsky83}.
Almost all glitching pulsars have a relatively low spin frequency
($\nus\lesssim 30\mbox{ Hz}$; see top panel of Figure~\ref{fig:glitch}),
such that they would have f-mode frequency corrections of $<1\%$.
There are two glitching pulsars with $\nus\approx327\mbox{ Hz}$,
but these have only been observed to glitch once and
are not expected to glitch again for a long time ($>100\mbox{ yr}$)
given their low spin-down rate, and each glitch was also very small in size,
i.e., $\Delta\nus\sim 10^{-9}\mbox{ Hz}$.
The only other fast-spinning glitching pulsar is PSR~J0537$-$6910 with
$\nus=62\mbox{ Hz}$ but is at a distant 50~kpc.
Thus based on current observational evidence, it appears safe to ignore
rotational effects.
On the other hand, we can see from the top panel of Figure~\ref{fig:ftau}
that a burst whose frequency is markedly distinct from the
distribution average could originate from a pulsar with
$\nus>50\mbox{ Hz}$.
Therefore detection of such a GW burst could indicate a fast-spinning
pulsar, especially since a (currently unknown) population of active
GW-emitting NSs may not share all the same properties as glitching pulsars.
It is also possible for f-modes of different spherical harmonic
to be excited.
The frequency difference between rotation-induced $m$=$0,\pm1,\pm2$
for $l$=2 modes is approximately the spin
frequency and would likely be resolvable for NSs with $\nus>1/T_{\rm obs}$,
where $T_{\rm obs}$ is time over which a GW search is performed.
The sub-second duration f-modes considered here imply
frequency splitting could be seen in bursts from pulsars with
$\nus>1\mbox{ Hz}$.

With a characteristic f-mode frequency $\nugw$ and damping time $\taugw$
assigned to each of the 188 known glitching pulsars, as well as their
measured spin frequency $\nus$ and distance $d$, and the glitch
size $\Delta\nus$ of each of the 552 measured glitches, we calculate
peak GW amplitude $\hamp$ using equation~(\ref{eq:hamp}).
Figure~\ref{fig:strain} shows the resulting peak GW strain
($=\hamp\sqrt{\taugw}$), as well as the spectral noise density
$\sqrt{S_h}$ of LIGO and next generation GW detectors
\cite{hildetal08,barsottietal18a,barsottietal18b,reitzeetal19}.
While the glitch size of most measured glitches and distance to
each corresponding pulsar produce
$\hamp\sqrt{\taugw}<10^{-24}\mbox{ Hz$^{-1/2}$}$,
about 20\% of glitches would be strong enough to produce a GW signal that is
observable by current and next generation detectors.
For example, bursts with the energy expected from a Vela glitch can reach
$\mbox{SNR}=\hamp\sqrt{\taugw/2S_h}\sim 5$ in advanced LIGO data
and $\sim 40$ using third generation detectors.
Bursts from a Crab-level glitch could have SNR $\sim 2$ using A+ and
$\sim 6$ using third generation detectors.

\paragraph{Discussion\label{sec:discuss}}
It is important to note that, since the true nuclear EOS is unknown
at this time, other model EOSs can yield average $\nugw$ and
$\taugw$ much lower or higher than the 2.4~kHz and 0.06~s obtained for
the BSk24 EOS, although their dispersions for a given EOS would be
similar to those shown in Figure~\ref{fig:ftau}.
Thus detection of bursts with average $\nugw$ significantly different
from 2.4~kHz does not invalidate our results but may indicate a
different EOS than the one considered here is preferred.
One can envision measuring bursts clustered around a particular
frequency due to f-mode oscillations (glitch-excited or by other means),
as well as burst signals at other frequencies due to entirely
different types of GW sources.
We should expect GW bursts from f-mode oscillations to obey the
$\nugw(M,R)$ and $\taugw(M.R)$ relations of
\cite{anderssonkokkotas96,anderssonkokkotas98,donevaetal13}
and $\nugw(M,I)$ and $\taugw(M,I)$ relations of
\cite{chirentietal15,donevakokkotas15}.
For example, bursts with higher $\nugw$ should have shorter $\taugw$.
Most should also have localization regions that overlap with the
Galactic plane.
An interesting avenue for future research is investigating data
analysis strategies based on an expected excess of transient
events in the relevant frequency range.

The GW strains shown in Figure~\ref{fig:strain} would seem to suggest
that GW bursts from systems like the Vela pulsar are essentially the only ones
that could be measured by current detectors, due to the pulsar's
proximity (287~pc) and large glitches
($\Delta\nus\gtrsim 10^{-5}\mbox{ Hz}$).
However, Vela glitches are relatively infrequent for GW searches,
occurring every 3--4~yr.
Thus one might expect the contribution of this type of burst source
to the total number of unmodeled transients detected by
LIGO/Virgo/KAGRA to be low.
However, our knowledge of the number of (nearby) glitching pulsars
and the number of glitches each pulsar undergoes is limited
because monitoring and timing each pulsar are crucial to
being able to measure glitches.
While there are only 15 known glitching pulsars at $< 1\mbox{ kpc}$,
there are actually more than 250 known pulsars at these distances
(see Figure~\ref{fig:psrdist}).
Some of these latter pulsars could have undergone
(electromagnetically unobserved) glitches and thus could contribute
to the number of GW bursts.
An advantage of GW observations is that they are not limited to
observing pulsars whose electromagnetic emission is beamed towards
us or that are electromagnetically-bright.
Thus there is potential for the type of GW source described here
to form a sizable fraction of transient signals detected by current
and future GW detectors.
It may even be possible to constrain the number of glitching pulsars
with GW data.
Finally it is important to reiterate that there is
currently no clear evidence for glitch-induced f-mode oscillations.
Nevertheless, these events provide a convenient known source population
with measured parameters and an illustration of the energies required to
produce detectable GW signals.

\begin{acknowledgments}
The authors appreciate the efforts of B. Shaw for maintaining the
Jodrell Bank Glitch Catalogue.
WCGH, DIJ, and NA acknowledge support through grant ST/R00045X/1 from
the Science and Technology Facilities Council in the United Kingdom.
CME acknowledges support from FONDECYT/Regular 1171421 and
USA1899-Vridei 041931SSSA-PAP (Universidad de Santiago de Chile).
\end{acknowledgments}

\bibliography{gwglitch}

\end{document}